\documentclass[conference, letter]{IEEEtran}
\ifCLASSINFOpdf
\else
\fi
\usepackage{comment}
\usepackage{multirow}
\usepackage{subcaption}
\usepackage{tabularx,booktabs}
\usepackage{graphicx}
\usepackage{color}
\usepackage{epstopdf}
\usepackage{setspace}
\usepackage{soul}
\usepackage{lettrine}
\usepackage[square,comma,sort&compress,numbers]{natbib}
\usepackage{amsmath}
\usepackage{amssymb}
\usepackage{algpseudocode,algorithm}
\graphicspath{{Figures/}}
\usepackage{multicol, blindtext}

\DeclareMathSizes{10}{9}{7}{6}
\IEEEaftertitletext{\vspace{-3\baselineskip}}

\begin{document}
\title {  Analysis and Optimization of HARQ for URLLC}  
\author{Faisal Nadeem,  Yonghui Li, Branka Vucetic, Mahyar Shirvanimoghaddam\\
School of Electrical and Information Engineering, The University of Sydney, NSW, Australia\\
Emails: \{faisal.nadeem, yonghui.li, branka.vucetic, mahyar.shm\}@sydney.edu.au}

\maketitle
\begin{abstract}
In this paper, we investigate the effectiveness of the hybrid automatic repeat request (HARQ) technique in providing high-reliability and low-latency in the finite blocklength (FBL) regime in a single user uplink scenario. We characterize the packet error rate (PER), throughput, and delay performance of chase combining HARQ (CC-HARQ) and incremental redundancy HARQ (IR-HARQ) in AWGN and Rayleigh fading channel with $m$ retransmissions. Furthermore, we consider a quasi-static fading channel model, which is more accurate than the over-simplified i.i.d. block fading or same channel assumption over consecutive packets. We use finite state Markov model under the FBL regime to model correlative fading. Numerical results present interesting insight into the reliability-latency trade-off of HARQ. Furthermore, we formulate an optimization problem to maximize the throughput performance of IR-HARQ by reducing excessive retransmission overhead for a target packet error performance under different SNRs, Doppler frequencies, and rate regimes.

% The new services of 5G i.e. Ultra reliable and low latency communication (URLLC) requires enhancement of reliability and latency simultaneously. In order to achieve less packet level latency, finite blocklength (FBL) regime is considered which decreases the reliability as short packet experience fewer channel realization. In order to compensate the loss in reliability and efficient utilization of resources,  retransmission techniques such as hybrid automatic repeat request (HARQ) can be used to increase packet length which achieve more time diversity of a channel. We optimize the retransmissions of HARQ for maximizing the PER and reducing the delay under the correlative fading channel. In this way controlled redundancy enables less latency.
\end{abstract}
\section{Introduction}
Cellular communications have primarily focused towards increasing spectral efficiency for enhance mobile broadband (eMBB) services. The fifth generation (5G) of mobile standard also includes ultra-reliable low-latency communication (URLLC) and massive machine-type communications (mMTC) into its area of focus \cite{holma20205g}. mMTC aims to support internet of things related application; therefore, enabling massive connectivity  of devices poses the main challenge \cite{shirvanimoghaddam2017fundamental}. The URLLC is another important direction as it will enable mission-critical services, such as remote surgery and wireless industrial control etc. URLLC is fundamentally challenging as it demands enhancement in two conflicting performance requirements, i.e., low-latency and high-reliability, simultaneously \cite{chen2018ultra}. In particular, for industrial control, a packet of size 32 bytes is required to be delivered with under 1ms latency and packet error rate (PER) of about $10^{-5}$, either with or without retransmission \cite{3GPPTS22261}.

In many communication scenarios, e.g., eMBB, usually packet arrival deadlines are not stringent and the packet size is large. Therefore, repetition of failing packets is a legitimate technique to achieve high throughput \cite{shafique2016cross}. Packet repetition with feedback is very effective in providing resilience against channel deep fades and distortions that arise anytime during transmission and efficient resource utilization \cite{shi2015analysis}. Therefore,  in latest communication standards, such as 4G/LTE and 5G, retransmission techniques like the hybrid automatic repeat request (HARQ) are adopted \cite{yang2018dynamic}. When a packet is successful with HARQ, the acknowledgment (ACK) signal is fed back to the transmitter, which sends a new packet; otherwise, it repeats the old packet when a negative ACK (NACK) is received.  HARQ has two common types, CC-HARQ and IR-HARQ known as chase combining and incremental redundancy HARQ,  respectively.  In CC-HARQ, the whole packet is repeated and maximum ratio combining (MRC) is used to increase the reliability. With IR-HARQ, the transmitter sends redundancy by sending a long codeword in chunks and the receiver improves reliability by increasing the decoding performance with a longer  codewords after each retransmission. HARQ increases reliability at the expense of increased latency  with each retransmission. Therefore, in the latest specifications of 5G, only 1 or 2 retransmissions of HARQ is supported \cite{3GPP_2018_PHY}. Both IR-HARQ  and CC-HARQ is studies in the finite blocklength (FBL) regime for its effectiveness in URLLC \cite{nadeem2021delay,shirvanimoghaddam2020dynamic}.

In addition to the signal-to-noise ratio (SNR) gains, HARQ  can also benefit from time diversity of the channel, when its retransmissions experience different channel gains due to relative mobility \cite{shi2015analysis}. In standards communication scenarios, when the packet duration is large, the channel is considered flat during the transmission of a packet and assumed to change independently during its retransmission \cite{shafique2016cross}. 
Whereas under the FBL regime, the channel is usually considered  fixed during the transmission and retransmission of a packet, which is an over simplification \cite{makki2019fast}.  A more accurate model to characterize the performance of HARQ is  the quasi-static fading  model \cite{sahin2019delay}. We analyse CC-HARQ and IR-HARQ in both AWGN and Rayleigh fading channel for its effectiveness for URLLC. We characterize the PER, throughput, and delay performance of each HARQ scheme. For the Rayleigh fading channel, we use the finite state Markov channel (FSMC) model  to characterize the effect of correlative fading on the HARQ performance. We also optimize the performance of HARQ in different channel conditions, average SNRs, code rates, and Doppler frequencies.  Finally, we propose an optimization framework for IR-HARQ to maximize its efficacy in providing the target level of reliability with lower latency.

The rest of the paper is organized as follows. The system model and preliminaries on HARQ in  the FBL regime are presented in Section \ref{Sec:System_Model}. In Section  \ref{SEC:HARQ_analysis} we present reliability, delay analysis, and optimization of HARQ. We provide numerical results in Section \ref{Sec:Numarical_resutls}. Finally, Section \ref{sec:conclusion} concludes the paper.

\section{System Model and Preliminaries}
\label{Sec:System_Model}
We consider a URLLC scenario that a stream of $N$ packets is transmitted and each $\ell$-th packet has its arrival deadline $T_{\ell}$. Let $s(t)$, $y(t)$, and $h(t)$ denote the transmitted signal, received signal, and channel gain at time $t$, respectively.  Then, $y(t)$ is given by:
\begin{align}
y(t)=h(t)s(t)+w(t),
\end{align}
where $w(t)\sim\mathcal{CN}(0,N_0)$ is the circularly symmetric zero-mean complex additive white Gaussian noise (AWGN). We also assume that the total transmit power is $\mathbb{E}[|s(t)|^2]=P_t$ and $s(t)$ is linearly modulated and transmitted with normalized symbol rate 1 symbol/s/Hz. We assume relative mobility between communicating terminals, which causes Doppler fading. The auto correlation function of $h(t)$ is modeled by a zeroth-order Bessel function of the first kind as $J_0(2\pi f_\mathrm{D}t)$, where $f_\mathrm{D}$ is the Doppler frequency.

\subsection{HARQ in the FBL regime}
We enable HARQ retransmissions to recover failing packets. When CC-HARQ is enabled, each failing packet is repeated in the next time slot and MRC is used to combine packets. For IR-HARQ, the transmitter encodes packets using  a $(n,k)$ channel code and sends initially  $n_1=\tau_1n$ symbols. If the packet is not successful (transmitter received a NACK), additional $n_2=\tau_2n$ symbols of the codewords are sent. This continues until the transmitter receives an ACK or maximum $n$ symbols are sent. 

The packet error rate for  CC-HARQ and IR-HARQ  with $(m-1)$ retransmissions in the FBL regime can be written as follows \cite{polyanskiy2010channel,9435801}\footnote{The authors in \cite{sahin2019delay} used a truncated systematic LT code to verify the normal approximation of standard HARQ i.e. \eqref{eq:CC-HARQ_AW} and \eqref{eq:IR-HARQ_AW}.  }
 \begin{align}
    {\epsilon}_\mathrm{cc}\left([\gamma_i]_1^m\right)\approx Q\left( \frac{n \log_2(1+\sum_{i=1}^m \gamma_i)-k+\log_2(n)}{n \sqrt{   V(\sum_{i=1}^m \gamma_i)}}\right),
    \label{eq:CC-HARQ_AW}
 \end{align}
\begin{align}
{\epsilon}_\mathrm{ir}&([\gamma_i]_1^m,[n_i]_1^m) \approx \nonumber \\
    & Q\left( \frac{\sum_{i=1}^m n_i \log_2(1+\gamma_i)-k+\log_2(\sum_{i=1}^m n_i)}{\sqrt{\sum_{i=1}^m  n_i V(\gamma_i)}}\right),
    \label{eq:IR-HARQ_AW}
 \end{align}
where $[x]_1^m=[x_1,\cdots,x_m]$, $V(\gamma_i)= \left(1-(1+\gamma_i)^{-2}\right) \log_2^2(e)$ is the channel dispersion \cite{polyanskiy2010channel}, $\gamma_i\in[\Gamma_{\ell}]_{1}^{L}$ is the SNR at the $i$-th round of HARQ packet transmission, and $Q(.)$ is the standard $Q$-function.

\subsection{Time block based FSMC }
\label{Sec: FSMC}

\begin{figure}[t]
\centering
\includegraphics[width=0.8\columnwidth]{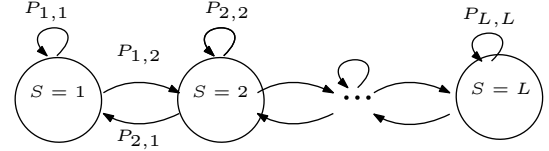}
\caption{The transport block based finite state Markov channel (FSMC) model of the Rayleigh fading channel.}
\label{fig:FSMM_Channel}
\vspace{-1ex}
\end{figure}

 In order to model correlative fading with finite  blocklength of a packet, we consider FSMC. In FSMC, we partition the fading envelop into $L$ fading states using thresholds $\boldsymbol\eta=[\eta_0,\eta_2,\ldots,\eta_{L}]$, where $\eta_1=0,\eta_{L+1}=\infty$. 
% Let $\mathcal{L}=[1, 2, \cdots, L]$ denote the index set for $L$ fading states, where $S_{\ell}$, for $\ell\in\mathcal{L}$, denotes the $\ell$-th fading state. 
At any given time, if the channel gain $|h(t)|$ lies in $[\eta_{\ell},\eta_{\ell+1}]$, the channel is said to be in state $S_{\ell}$.
We set $\eta_\ell$ for \emph{equal duration partitioning} of the fading envelop so that $n$ length packet experience single fading state during its transmission \cite{zhang1999finite}. This forms a block by block  discrete time channel variation model and form a Markov chain, that is sampled at each time block (TB) shown in Fig. \ref{fig:FSMM_Channel}. The fading envelope partition $\eta_\ell$, the TB duration $t_{\text{TB}}$, and the Doppler frequency $f_D$ are sufficient to fully characterize the underlying Markov chain. Under the FBL assumption, we simplify the FSMC by assuming   $P_{\ell,k}=0$ whenever $|\ell-k|>1$, i.e. a state can only transitions to adjacent states, where $P_{\ell,k}$ is the transition probability from channel fading state ${\ell}$ to state ${k}$, where $\ell,k\in \mathcal{L}$. This is a valid assumption,  because in the FBL regime the channel variation is low across consecutive packets.
% Fig. \ref{fig:FSMM_Channel} shows the state transition diagram of the finite state MM for fading channel. 
\begin{figure}[t]
\centering
\includegraphics[width=\columnwidth]{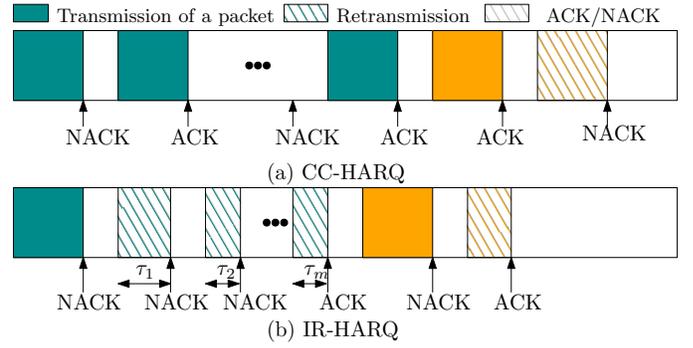}
\caption{The packet transmission in IR-HARQ with $m$ retransmissions}
\label{fig:M_N-HARQ}
\vspace{-1ex}
\end{figure}
The state transition probabilities associated with the FSMC model can be calculated as follows
\cite{wang1995finite}:
\begin{subequations}
\begin{align} 
P_{\ell,{\ell}+1}\left ({\boldsymbol\eta, f_{\mathrm{D}}, t_{\mathrm{TB}}}\right)&\approx\frac {N(\eta _{\ell+1}, f_{\mathrm{D}})t_{\mathrm{TB}}}{q_{\ell}(\eta_{\ell},\eta_{\ell+1})},\quad 1 \le\ell\le L-1, \label{eq:1}\\ 
P_{\ell,{\ell}-1}\left ({\boldsymbol\eta, f_{\mathrm{D}}, t_{\mathrm{TB}}}\right)&\approx\frac {N(\eta_{\ell}, f_{\mathrm{D}})t_{\mathrm{TB}}}{q_{\ell}(\eta_{\ell},\eta_{\ell+1})},\quad ~~~2\le\ell\leq L\label{eq:2}
\end{align}
\label{eq:fadingtransitions}
\end{subequations}
where $q_\ell$ is the marginal probability of channel being in ${\ell}$-th state given as:
\begin{align} 
q_{\ell}= \int_{\eta_{\ell}}^{\eta_{\ell+1}} {2 x e^{-x ^{2}} dx} ,
\label{eq:marg_prob_states}
\end{align}
where $f(x)=2 x e^{-x^2}$ is the  probability distribution of $|h(t)|$ and under the Bessel auto-correlation function $N(\eta_{\ell},f_D)$ is the average number of times per second, the signal envelope $|h(t)|$ crosses level $\eta_{\ell}$  given by \cite{wang1995finite,sadeghi2008finite}:
\begin{align} 
N(\eta_{\ell},f_{\mathrm{D}})=\sqrt{2\pi}\eta_{\ell} f_{\mathrm{D}} e^{-\eta_{\ell}^{2}}.
\label{eq:channelcrossing}
\end{align}

Equation \eqref{eq:fadingtransitions} and \eqref{eq:channelcrossing} indicate that state transition probabilities linearly increase with $t_\mathrm{TB}$ and $f_D$. Therefore,  for  total outgoing probabilities of each state to be 1, $t_\mathrm{TB}$  must be upper bounded as
\begin{align}
  t_\mathrm{TB}\leq\frac{p_\ell(\eta_\ell, \eta_{\ell+1})}{N(\eta_{\ell}, f_{\mathrm{D}})+N(\eta_{\ell+1}, f_{\mathrm{D}})},\quad \forall \ell\in \mathcal{L},
  \label{eq:state_duraton}
\end{align}
where \eqref{eq:state_duraton} indicates that $t_\mathrm{TB}$ cannot exceed the average duration of the state. Following this, we can set the blocklength $n(B,t_{\mathrm{TB}})=B t_{\mathrm{TB}}$. Note that thresholds $\mathbf{\eta}$ are uniquely specified to provide equal state duration based on  $f_\mathrm{D}$ and $t_\mathrm{TB}$ values  \cite{zhang1999finite}. 
With crossover probabilities modeled by AWGN, the symbol level SNR when the channel gain is $|h(t)|=x$, is given by $\gamma(x,t)=\frac{P_{\mathrm{t}} x^2}{B N_0}$. The normalized SNR at the $\ell^{th}$ channel state, denoted by $\Gamma_{\ell}$  is then given by \cite{sahin2019delay}:
\begin{align} 
\Gamma_{\ell}&= \frac {\int_{\eta_{\ell}}^{\eta_{\ell+1}} { \gamma(x,t) 2 x e^{-x^{2}} dx }}{q_{\ell}( \eta_\ell,\eta_{\ell+1})} \nonumber \\
&=\frac{P_{\mathrm{t}}}{BN_0} \frac {e^{-\eta_\ell^2}(\eta_\ell^2+1)-e^{-\eta_{\ell+1}^2}(\eta_{\ell+1}^2+1)}{e^{-\eta_\ell^2}-e^{-\eta_{\ell+1}^2}}.
\label{eq:SNR_soft2}
\end{align}

\section{Reliability and delay analysis}
\label{SEC:HARQ_analysis}
\subsection{AWGN Channel}
Let $\mathcal{A}_{i}$ denote the events that the decoding failed at the receiver after $i$ retransmissions. Then, the probability that a packet is decoded successfully after exactly $i$ retransmissions, denoted as $p_i$ for $i\in\{1,\cdots,m\}$, is given by
 \begin{align}
    \nonumber p_{i}&=\mathrm{Prob}\left\{\mathcal{A}_{i-1} \bigcap \mathcal{A}^{c}_{i}\right\}\\
    \nonumber
    &=\mathrm{Prob}\{\mathcal{A}_{i-1}\}-\mathrm{Prob}\left\{ \mathcal{A}_{i-1}\bigcap \mathcal{A}_{i}\right\}\\
    \nonumber &\overset{(a)}{\approx} \mathrm{Prob}\{\mathcal{A}_{i-1}\}-\mathrm{Prob}\left\{\mathcal{A}_{i}\right\}\\
    &=\epsilon\left([\gamma]_1^{i-1}, [n_i]_1^{i-1}\right)-\epsilon\left([\gamma]_1^{i},[n_i]_1^{i}\right),
\end{align}
where $n_i=\tau_i n$ and step $(a)$ follows from the fact that the decoding with $i-1$ packet almost certainly fails if the decoding failed with $i$ packets. We also have packet success probability with single transmission as
$p_0=1-\epsilon([\gamma_0],[n_0])$, and $p_e=\epsilon\left([\gamma]_1^{m},[n_i]_1^{m}\right)$ is the error probability, where $\tau_0=1$ and $\epsilon(.)$ is the error rate in the finite blocklength regime which is given in \eqref{eq:IR-HARQ_AW} and  \eqref{eq:CC-HARQ_AW} for IR-HARQ and CC-HARQ, respectively.  For CC-HARQ, we have $\tau_i=1$ as all retransmission packets have the same length as the original packet.%Also  $\gamma_i=\gamma_0=P/N_0$ is the SNR in each transmission. 

\subsection{Rayleigh fading Channel}
 In order to characterize the PER of IR-HARQ and CC-HARQ in the Rayleigh fading channel, we assume a finite state Markov channel model (FSMC) \cite{sadeghi2008finite}. Similar to \cite{sahin2019delay}, the FSMC model is used under FBL assumption. In the FSMC, the overall fading envelop is partitioned into $L$ fading states where SNR of $\ell$-th fading state denoted as $\Gamma_\ell$.
 Then the success and fail probabilities of a packet when $m=2$  can be calculated directly using marginal  probabilities of each $\ell$-th fading state and transitioning probabilities $P_{\ell,k}$ as follows:
\begin{align}
    p_0=&\sum_{\ell\in\mathcal{L}}q_{\ell}\left(1-\epsilon\left([\Gamma_\ell],[n_0]\right)\right)\\ 
     p_1=&\sum_{\ell\in\mathcal{L}}\sum_{k\in\mathcal{L}}q_{\ell}P_{\ell,k}(\epsilon([\Gamma_\ell],[n_0])-\epsilon([\Gamma_\ell,\Gamma_k],[n_0,n_1]))\\ 
     p_e=&\sum_{\ell\in\mathcal{L}}\sum_{k\in\mathcal{L}}q_{\ell}P_{\ell,k}\epsilon([\Gamma_\ell,\Gamma_k],[n_0,n_1])
\end{align}
where $q_\ell$ is the marginal probability of fading $\ell$-th state and $P_{\ell,k}$ is the transition probability between fading state $\ell$ and $k$
given in detailed in Section \ref{Sec: FSMC}.

It is then easy to show that the throughput for HARQ with $m-1$ retransmissions is given by:
\begin{equation}
  \eta= \frac{k}{n} \times \frac{1-p_e}{ p_0 + \sum_{i=1}^{m-1}\big(p_{i}\sum_{j=0}^{i}\tau_{j} \big) +   p_{e}\sum_{i=0}^{m-1} \tau_{i} }.
  \label{eq:throughput_AWGN}
\end{equation}
The delay distribution for a given packet with HARQ is given by:
\begin{align}
    D[d]=p_0\delta[d-1]&+\sum_{i=1}^{m-1}p_{i}\delta\left[d-\sum_{j=0}^i\tau_{j}\right]+p_e\delta\left[d-\sum_{i=0}^{m-1}\tau_{i}\right]
    \label{eq:Delay_distrm}
\end{align}
where $\delta[d]$ is the discrete Dirac delta function. It is then easy to show that the overall delay distribution for delivering $N$ packets (either successful or unsuccessful) with HARQ can be found as follows:
\begin{align}
D^{(N)}_\text{O}[d]=\bigotimes_{i=1}^ND[d]
\end{align}
where $\otimes$ is the convolution operand. For example, when $m=2$, the delay distribution of delivering $N$ packets using CC-HARQ is given by \cite{nadeem2020non}:
\begin{align}
    D^{(N)}_\text{O}[d]=\sum_{i=0}^{N}\dbinom{N}{i}(1-\epsilon([\gamma_0],[n]))^{i}&\epsilon([\gamma_0],[n])^{N-i} \nonumber &\\\times\delta[d-(1+\tau_1)N+i],
    \label{eq:Qdelay_N1}
\end{align}

\begin{figure*}[htb]
    \centering
    \begin{subfigure}{.495\textwidth}
    \includegraphics[width=\columnwidth]{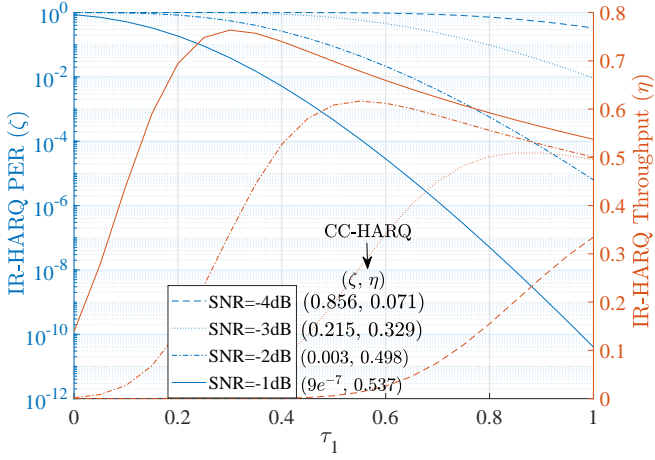}
   \caption{$k=100$}
    \label{fig:PERtpSNR} 
    \end{subfigure}
    \begin{subfigure}{.495\textwidth}
    \centering
    \includegraphics[width=\columnwidth]{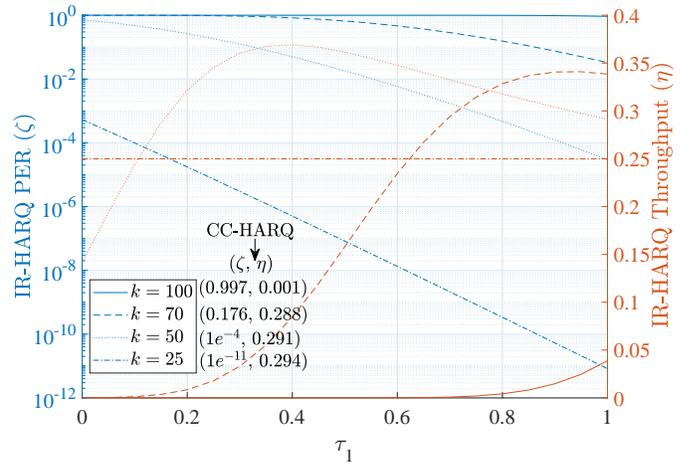}
    \caption{  SNR$=-5dB$ }
    \label{fig:PERtpk}
    \end{subfigure}
    \caption{Variation of PER and throughput with $\tau_1$ at various SNRs and $k$ for IR-HARQ and CC-HARQ in the AWGN channel when packet length $n=100$.}
    \vspace{-3ex}
    \label{fig:perthpjoint}
\end{figure*}
\section{Numerical Results}
\label{Sec:Numarical_resutls}
In this section, we present simulation results to highlight the reliability and delay performance of IR-HARQ  and CC-HARQ under AWGN and Rayleigh fading channel. The optimization of IR-HARQ is also presented to improve performance further. 

\subsection{Performance in AWGN channel}
Fig. \ref{fig:perthpjoint} shows the PER and throughput performance of CC-HARQ and IR-HARQ with different SNRs and rate $R(n,k)$. We vary rate by changing $k$ and keeping $n$ fixed. In CC-HARQ, retransmission is conducted by repeating the entire codeword when requested. For IR-HARQ, PER and throughput versus retransmission coefficient $\tau_1$ is presented.  In general, the PER performance of IR-HARQ and CC-HARQ improves by increasing SNRs and reducing rate, i.e. lowering $k$. The PER improves due to  retransmission of failing packet, however, throughput gain starts to saturate for IR-HARQ as the retransmission redundancy increases beyond requirement. For CC-HARQ as the entire codeword is retransmitted; therefore, throughput is not always maximum at a given SNR or $k$ as can be seen in  Fig. \ref{fig:perthpjoint}. Because the failing packet sometimes can recover with little redundancy.  More specifically, in Fig. \ref{fig:PERtpSNR} for target PER $10^{-4}$ with IR-HARQ, $\tau_1=0.58$ is required at SNR=-1dB in order to achieve highest throughput. When SNR is less, e.g. SNR=-2dB, more retransmission is required to achieve PER $10^{-4}$, e.g. $\tau_1=1$. On the other hand, with CC-HARQ at an SNR=-1dB, the throughput is 0.537 with excessive redundancy.
 
\begin{figure}[t]
    \centering
    \includegraphics[width=\columnwidth]{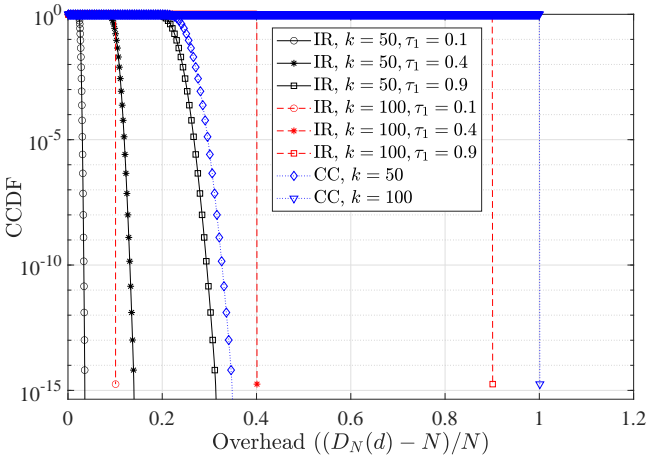}
    \caption{Delay overhead distribution of CC-HARQ and IR-HARQ at SNR=-4dB and $n=100$ with different rates $R$. }
    \label{fig:Del_CCDF} 
\end{figure} 

Fig. \ref{fig:Del_CCDF} shows the delay performance of IR-HARQ and CC-HARQ.  We assume that $N=1000$ packets are scheduled for transmission and each transmission occupies a sing time slot.  Since  retransmissions requires additional time slot, we normalize the delay of 1000 packets in  Fig. \ref{fig:Del_CCDF}.  Therefore, when all the packets are delivered with single transmission the delay overhead is zero. Any additional delay due to retransmission is counted as delay overhead.  Figure   \ref{fig:Del_CCDF} presents the distribution of the delay overhead. As can be seen in Fig. \ref{fig:Del_CCDF}, when rate ($k$) is higher, more packets require retransmission leading to higher delay overhead. Furthermore, each retransmission delays the new arriving packets, leading to queuing and larger delay overheads. 
A low PER setting, e.g. a small $k$, reduces the queuing delay; however, this leads to low throughput as shown in Fig \ref{fig:PERtpk}.
Note that smaller retransmission coefficients $\tau_1$ can significantly limit the queuing delay.  Because, a shorter retransmission reduces the queuing delay of the subsequent packets.  Consequently at for a target PER and throughput, minimizing $\tau_1$  leads to less queuing delay. For example,  at  $k=50$, $\tau_1=0.4$ achieves the desired level of reliability with $\approx 20\%$ less packets delay overhead than $\tau_1=0.9$.

\begin{figure*}[t]
    \centering
    \begin{subfigure}{.495\textwidth}
    \includegraphics[width=\columnwidth]{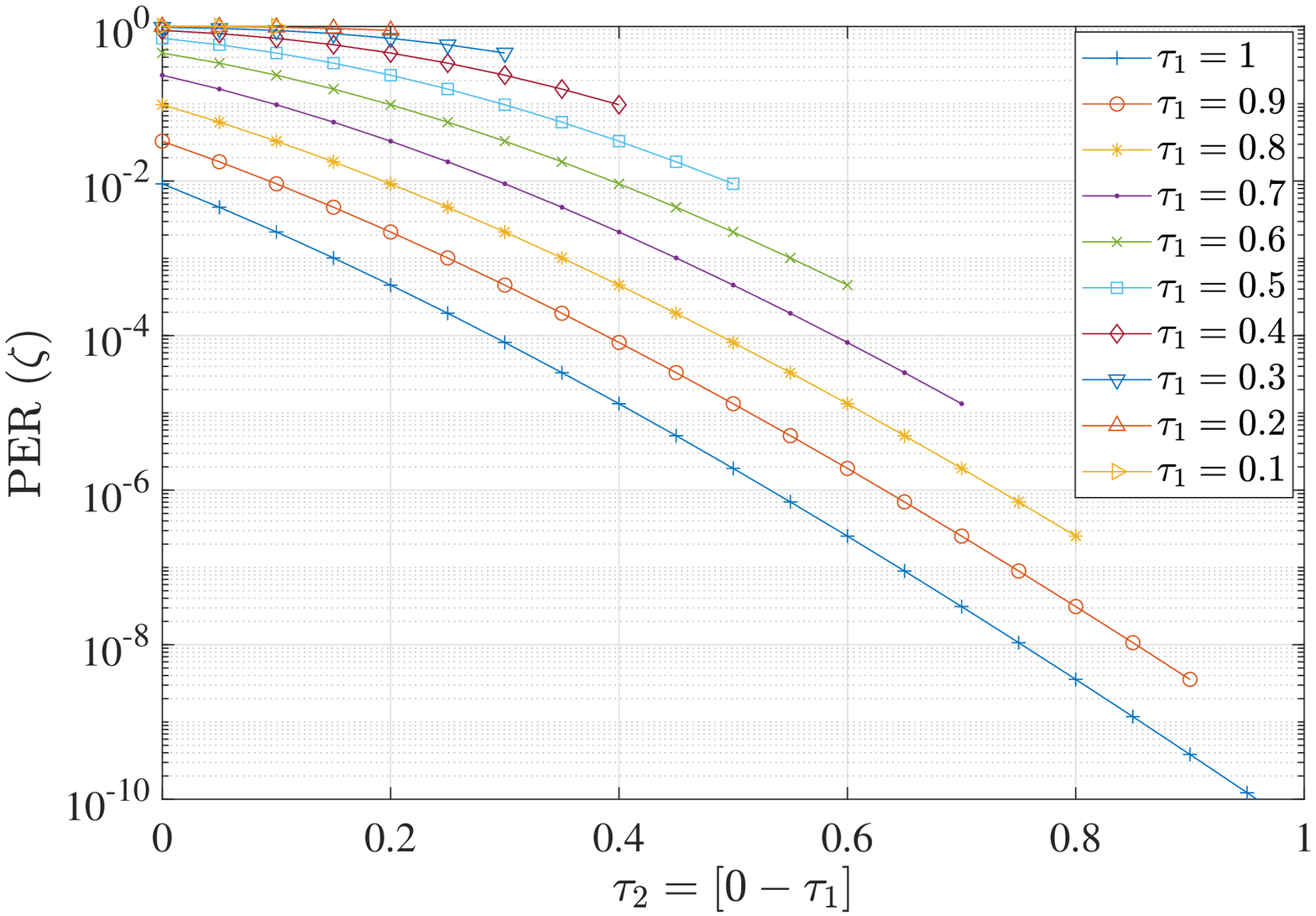}
    \caption{PER}
    \label{fig:PERm3} 
    \end{subfigure}
    \begin{subfigure}{.495\textwidth}
    \centering
    \includegraphics[width=\columnwidth]{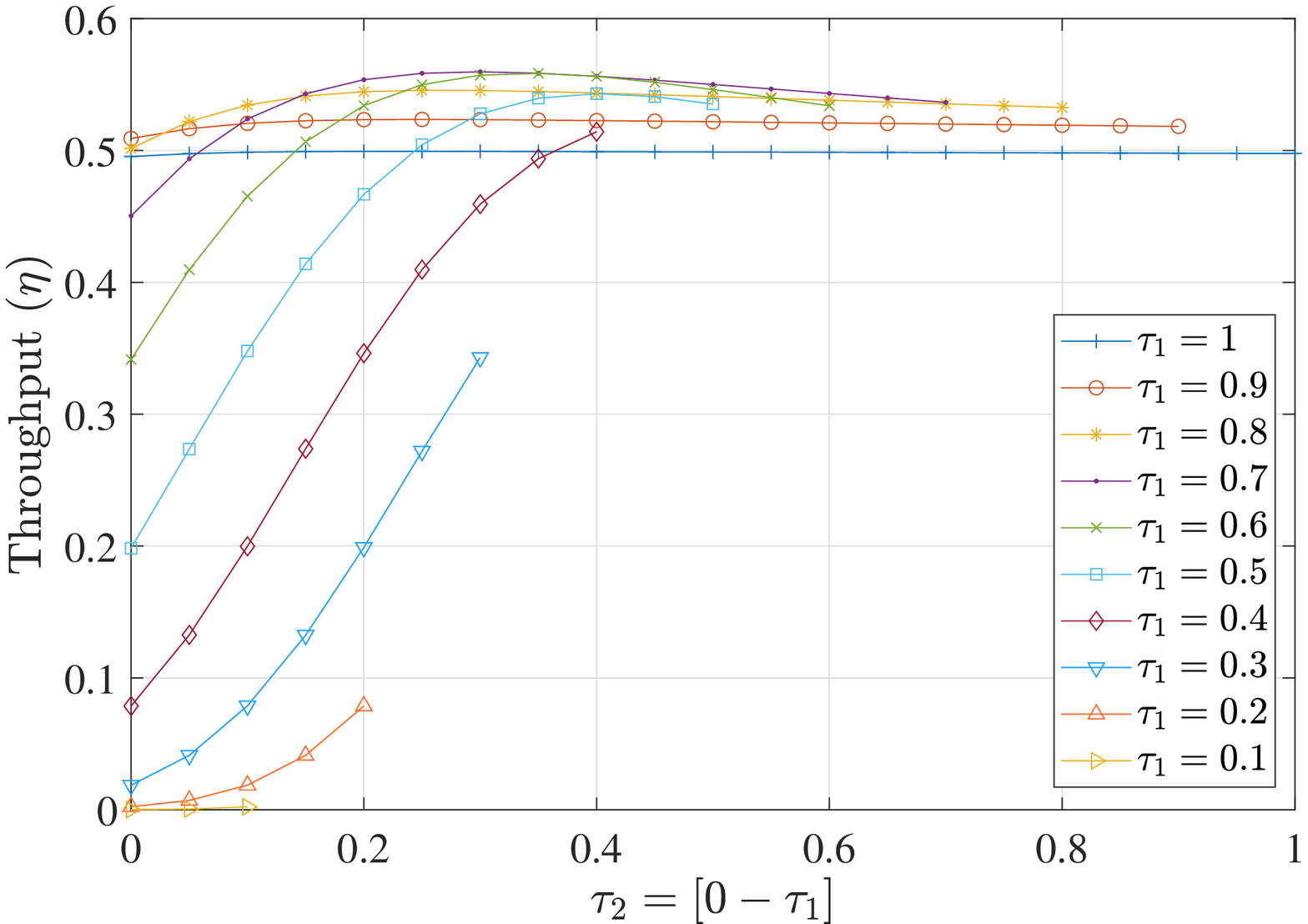}
    \caption{Throughput}
    \label{fig:THpm3}
    \end{subfigure}
      \caption{PER and throughput performance of IR-N-HARQ in the AWGN channel when $m=3$, $k=70$, $n=100$, SNR$=-4$dB. }
      \vspace{-3ex}
    \label{fig:perthpntm3}
\end{figure*}

Fig. \ref{fig:perthpntm3} shows PER and Throughput performance of IR-HARQ in FBL, when two retransmissions are enabled. For practical consideration, the incremental redundancy increase with each retransmission; therefore,  we set  $\tau_2\leq\tau_1$, where $\tau_2$ is the fraction of redundancy in second retransmission.  This figure shows that with proper retransmission parameters ($\tau_1,\tau_2$), IR-HARQ can achieve a higher reliability. More specifically, with $\tau_1=\tau_2=1$ the maximum redundancy is retransmitted, which corresponds to the lowest PER  as can be seen in Fig. \ref{fig:PERm3}. However, Fig. \ref{fig:THpm3} shows that at a fixed SNR, choosing appropriate $\tau_1$ and $\tau_2$ increases throughput with a certain level of PER performance. For example, with $\tau_1=0.7$ and $\tau_2=0.6$, PER$~10^{-4}$ can be achieved with higher throughput than $\tau_1=\tau_2=1$. Furthermore, with smaller retransmission coefficients, the overall latency and queuing delay is also reduced as can be seen in Fig. \ref{fig:Del_CCDF} for $m=2$.

\subsection{Performance in Rayleigh fading channel}
\begin{figure*}[t]
    \centering
    \begin{subfigure}{.495\textwidth}
    \includegraphics[width=\columnwidth]{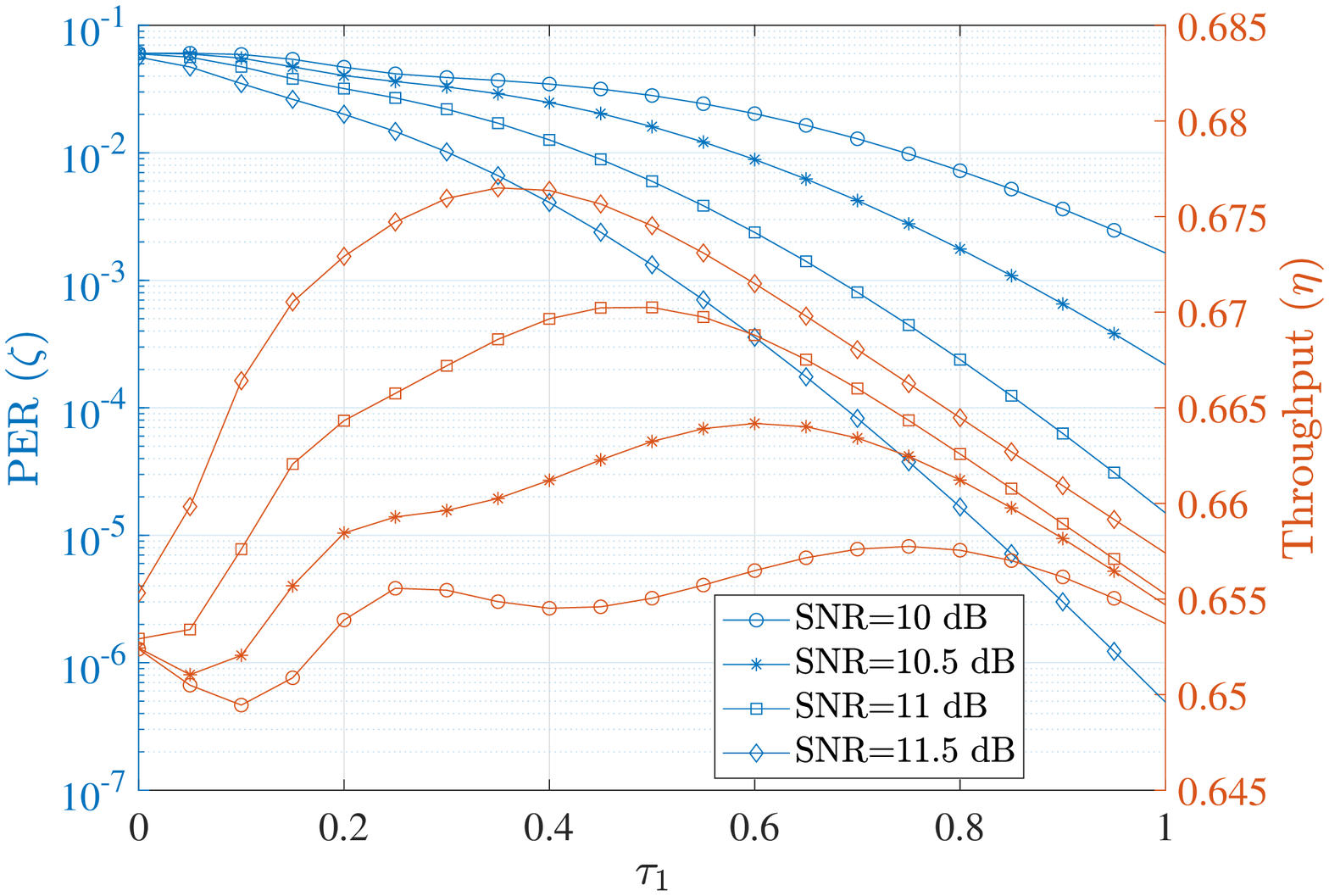}
    \caption{$k=70$}
    \label{fig:fdk1} 
    \end{subfigure}
    \begin{subfigure}{.495\textwidth}
    \centering
    \includegraphics[width=\columnwidth]{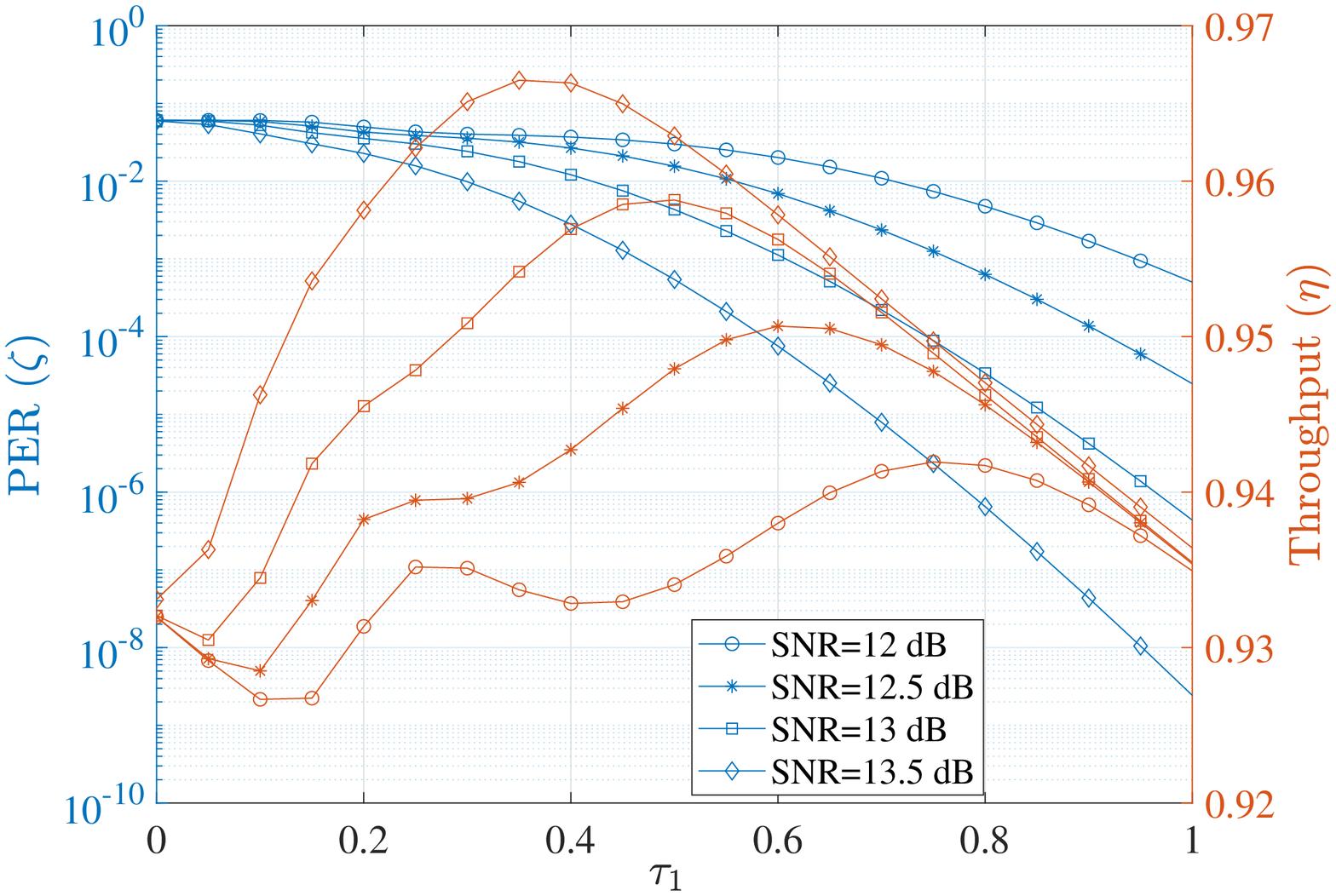}
      \caption{$k=100$}
    \label{fig:fdk2}
    \end{subfigure}
      \caption{PER and throughput performance of IR-HARQ with various  $\tau_1$ at $n=100$, $t_\mathrm{TB}=0.00014$ $f_\mathrm{D}t_\mathrm{TB}=0.0338$.  }
      \vspace{-3ex}
    \label{fig_fdall}
\end{figure*}
We utilize FSMC and  model the quasi static fading by splitting the fading envelop into $L$ fading states. The boundaries of each fading states are chosen to model equal duration of fading states called equal duration channel partitioning. More specifically, assume $c$ as a channel partitioning parameter representing the number of  packets experiencing a specific fading state. The choice of $c$ is usually between 1 and 8 \cite{sahin2019delay}. We set $c=3.0446$. The   channel variability across packets depends upon the length of the packet named as time block duration $t_\mathrm{TB}$ i.e. $f_\mathrm{D}t_\mathrm{TB}$ value indicates the relative fading speed.  In low mobility, i.e. small $f_\mathrm{D}t_\mathrm{TB}$, total number of fading states $L$ needs to be incensed for any fixed $c$ to capture  the  fading variation across consecutive packets more accurately \cite{zhang1999finite}.   We study low mobility and high mobility cases with the following two sets of parameters. When $f_\mathrm{D}=210$Hz, and  $t_\mathrm{TB}=0.14$ms, we choose $L=13$ and when $f_\mathrm{D}=285$Hz, $t_\mathrm{TB}=0.3$ms we choose $L=4$. With carrier frequency $f_\mathrm{C}=1.9$ GHz, these setting capture relative velocity range of $20-80$ Km/hr.

Fig. \ref{fig_fdall} shows the PER and throughput performance of IR-HARQ  with various $\tau_1$ at different SNRs and information rate $R$. Similar to AWGN case, overall PER  reduces linearly with higher retransmission coefficient in Rayleigh fading case and throughput is saturating due to excessive redundancy.  Also, the trend of PER and throughput curves show that there exist trade-off between the achievable PER and throughput. For example  in Fig. \ref{fig:fdk1}  when $k=70$ , by setting $\tau_1=0.4$ the throughput is maximized at 0.675 bits/sec/Hzm when SNR 11.5dB. This parameter setting corresponds to PER $\sim 10^{-2.5}$. The PER can be further reduced to $10^{-4}$ by increasing $\tau$ to 0.6,  which slightly reduces the throughput  to 0.625.  Fig. \ref{fig:fdk2} shows that  increasing information length $k$ and  SNRs with proper choice of $\tau_1$  further improves throughput performance.

\subsection{Optimization of retransmission coefficients}
With IR-HARQ  the retransmission coefficient can be chosen to  maximize the throughput for target PER that also results in less queuing delay due to optimal retransmission length. 
For a given SNR  and rate $R(n,k)$, we have the following optimization problem when $m=2$
\begin{align}
\label{eq:opt}
&\max_{\tau_1} \; \eta \\
\nonumber \textrm{s.t.}\;\; 
&\mathrm{C}_1.\; 0<\tau_1\leq 1, \;\nonumber\\ 
&\mathrm{C}_2. \;\zeta \geq\zeta_0,\nonumber
\end{align}
where  $\mathrm{C}_1$ is the range of retransmission coefficient  and $\mathrm{C}_2$ implies that the PER should not exceed certain threshold  $\zeta_0$.

\begin{table*}[t]
\caption{Optimum values of $\tau_1$ for IR-HARQ at $n=100$ for various SNRs and rate $R(n,k)$.}
\label{TabOpt}
\begin{subtable}{0.49\textwidth}
%  \begin{subtable}{1\columnwidth}
\setlength{\tabcolsep}{13pt}
 \centering
\caption{$k=100$}
\label{Table2}
\scriptsize
 \begin{tabular}{ccccc}
\toprule
 &   \multicolumn{2}{c}{$f_\mathrm{D}t_\mathrm{TB}=0.0338$}    
 &   \multicolumn{2}{c}{$f_\mathrm{D}t_\mathrm{TB}=0.04$}    \\ 
  \cmidrule{2-3} \cmidrule{4-5} 
SNR & $\hat{\tau}_1$, & $(\zeta, \eta)$ & $\hat{\tau}_1$ & $(\zeta, \eta)$ \\
 \midrule
11  & 0 &    0.0605,  0.9319 &  0.7&  0.0055, 0.9330   \\ 
11.5 & 0.9 &  0.0089, 0.9323 & 0.5  &  0.0093, 0.9440 \\ 
\bf{12}   & 0.8 &  0.0047, 0.9417& \bf{0.4}  &\bf{0.0064}, \bf{0.9546} \\ 
\bf{12.5} & \bf{0.6}&  \bf{0.0068},  \bf{0.9506}& 0.3  &   0.0047, 0.9652 \\ 
13 & 0.5 &0.0043, 0.9587& 0.2 & 0.0040,  0.9767  \\ 
13.5   & 0.4&0.0027,  0.9663& 0.2 & 0.0004,   0.9859\\ 
14 & 0.3& 0.0019, 0.9745& 0.1  & 0.0004, 0.9908\\ 
\bottomrule
 \end{tabular}
 \end{subtable}
\begin{subtable}{0.49\textwidth}
\setlength{\tabcolsep}{13pt}
\centering
\caption{$k=70$}
\label{Table1}
\scriptsize
 \begin{tabular}{ccccc}
\toprule
 &   \multicolumn{2}{c}{$f_\mathrm{D}t_\mathrm{TB}=0.0338$}    
 &   \multicolumn{2}{c}{$f_\mathrm{D}t_\mathrm{TB}=0.04$}    \\
  \cmidrule{2-3} \cmidrule{4-5} 
SNR & $\hat{\tau}_1$ & $(\zeta, \eta)$ &  $\hat{\tau}_1$ & $(\zeta, \eta)$ \\
 \midrule
11   & 0.5 & 0.0059, 0.6702 & 0.2& 0.0049, 0.6845  \\ 
11.5 & 0.4 & 0.0040,  0.6763  & 0.2  &  0.0008, 0.6906  \\ 
12   & 0.3 &0.0029,  0.6829&  0.1  & 0.0007, 0.6934  \\ 
\bf{12.5} & 0.2& 0.0022, 0.6888& \bf{0.1}  & $\bf{6.5e^{-5}}$, \bf{0.6943}\\ 
13   &0.2& 0.0003, 0.6925& 0.1 & $2.7e^{-6}$, 0.6944  \\ 
13.5   &  0.1& 0.0002,  0.6940& 0.1  & $5.0e^{-8}$, 0.6944 \\ 
\bf{14}   &\bf{0.1}& $\bf{1.9e^{-5}}$, \bf{0.6944}& 0.1  & $3.6e^{-10}$, 0.6944 \\ 
\bottomrule
 \end{tabular}
 \end{subtable}
\end{table*}

Table \ref{TabOpt} presents the optimization of retransmission coefficient $\tau_1$, when $k=100$ and $70$ with various SNRs to achieve various PER reliability and throughput.  The optimized parameters are  provided in Table I for the Rayleigh fading case and two different relative mobility scenarios, i.e $f_\mathrm{D}t_{\mathrm{TB}}=0.0338$ and $0.04$. Firstly, it can be seen that throughput is maximized by keeping PER at the desired level with $\tau_1$ smaller than $1$. Furthermore, as SNR increases, $\tau_1$ can be further reduced while maintaining roughly similar PER and throughput performance. For example, in Table \ref{TabOpt}(a), at $f_\mathrm{D}t_{\mathrm{TB}}=0.0338$,  $\zeta~0.0047$ and $\eta~0.94$ is achieved at SNR $12$dB, with $\tau_1=0.8$, which  reduces to $\tau_1=0.6$  with additional $0.5$dB  SNR for similar reliability.  Secondly,  when relative mobility increases, $\tau_1$ can be further decreased without loosing the PER and throughput performance at a given SNR. More specifically, in Table \ref{TabOpt}(a),  similar $(\zeta,\eta)$ performance is achieved with $\tau_1=0.4$ when mobility is high i.e. $f_\mathrm{D}t_{\mathrm{TB}}=0.04$, whereas when at low mobility, i.e. $f_\mathrm{D}t_{\mathrm{TB}}=0.0338$ $\tau_1=0.8$. This is primarily due to the fact that with higher relative mobility channel variation between first transmission and its retransmission is higher which provides more diversity and reliability. Therefor, less retransmission is required when mobility is high for a target PER and throughput at fixed SNR or rate. Note that smaller values of $\tau_1$ also leads to less queuing delay as shown in Fig. \ref{fig:Del_CCDF}.  Table I can be used to select the optimal $\tau_1$ according to fading parameters to avoid excess SNR.
% In other words for fixed $\tau$ SNR gain can be observed when relative mobility is increased.
For example in \ref{TabOpt}(a), the SNR gain is $0.5$dB for target $\zeta~0.006$ and $\eta~0.95$. Also as shown in Table \ref{TabOpt}(b), the SNR gap is ~2.5 dB  at $\tau_1=0.1$ for similar PER and throughput target of ~$10^{-5}$ and $0.694$, respectively.

\section{Conclusion}
\label{sec:conclusion}

In this paper, we analyzed packet retransmission techniques, i.e. IR-HARQ and CC-HARQ, in the FBL regime for URLLC. More specifically, PER, throughput and delay performance is characterized analytically under AWGN and Rayleigh fading channel allowing $m$ retransmissions. Finite state Markov channel  model is used to represent quasi-static fading  under short packet communication for accurate representation of diversity gain with retransmission.  Simulations are carried out to get further insight into the reliability and delay performance of HARQ schemes. Finally, we formulated an optimization problem to maximize throughput for target PER performance under various SNRs and Doppler frequencies. The optimization of retransmission coefficient also translated to the  queuing delay minimization. 

%   the usefulness of optimizing  retransmission coefficient with various  for throughput maximization under target PER performance.

% \hl{In this paper, we analyzed and optimized HARQ for URLLC in both AWGN and Rayleigh fading c}hannel. The PER throughput and delay is characterized under correlative fading instead of assuming i.i.d block fading model.  Using optimal retransmission length of IR-HARQ especially in relation to correlative fading, the delay is reduced significantly. We also demonstrate using numerical simulations, the optimization of retransmission for $m=3$.  The simulations results demonstrated that optimization of retransmission is useful in reducing the overall delay.  
% \appendices
\footnotesize
\bibliographystyle{IEEEtran}
\bibliography{myReferences}

\end{document}